\documentclass{aastex63}
\usepackage{xspace,graphics,graphicx,hyperref}
\newcommand\teff{\ensuremath{T_\mathrm{eff}}\xspace}
\newcommand\logg{\ensuremath{\log g}\xspace}
\newcommand\msun{\ensuremath{M_\odot}\xspace}
\accepted{for Publication in the Astronomical Journal January 29, 2021 }

\shorttitle{M67 White Dwarf Properties}
\shortauthors{Canton et al.}

\turnoffedit

\begin{document}

\title{The White Dwarfs of the Old, Solar Metallicity Open Star Cluster Messier 67: Properties and Progenitors \footnote{Some of the data presented herein were obtained at the W.M. Keck Observatory, which is operated as a scientific partnership among the California Institute of Technology, the University of California and the National Aeronautics and Space Administration. The Observatory was made possible by the generous financial support of the W.M. Keck Foundation.}}

\author[0000-0002-2839-1766]{Paul A. Canton}
\affiliation{Homer L.~Dodge Department of Physics and Astronomy, University of Oklahoma, 440 W.~Brooks St., Norman, OK 73019, USA}

\author{Kurtis A. Williams}
\affiliation{Department of Physics \& Astronomy, Texas A\&M University-Commerce, P.O. Box 3011, Commerce, TX, 75429-3011, USA}

\author[0000-0001-6098-2235]{Mukremin Kilic}
\affiliation{Homer L.~Dodge Department of Physics and Astronomy, University of Oklahoma, 440 W.~Brooks St., Norman, OK 73019, USA}

\author{Michael Bolte}
\affiliation{UCO/Lick Observatory, University of California, 1156 High St., Santa Cruz, CA 95064, USA}

\defcitealias{2018ApJ...867...62W}{Paper~I}

\begin{abstract}
The old, solar metallicity open cluster Messier 67 has long been considered a lynchpin in the study and understanding of the structure and evolution of solar-type stars.  The same is arguably true for stellar remnants -- the white dwarf population of M67 provides crucial observational data for understanding and interpreting white dwarf populations and evolution. In this work, we determine the white dwarf masses and derive their progenitor star masses using high signal-to-noise spectroscopy of warm ($\gtrsim10,000$ K) DA white dwarfs in the cluster. From this we are able to derive each white dwarf's position on the initial-final mass relation, with an average $M_{\mathrm WD} = 0.60\pm 0.01 \msun$ and progenitor mass $M_i = 1.52\pm 0.04\msun$.  These values are fully consistent with recently published linear and piecewise linear fits to the semi-empirical initial-final mass relation and provide a crucial, precise anchor point for the initial-final mass relation for solar-metallicity, low-mass stars.  The mean mass of M67 white dwarfs is also consistent with the sharp narrow peak in the local field white dwarf mass distribution, indicating that a majority of recently-formed field white dwarfs come from stars with progenitor masses of $\approx 1.5~\msun$.  Our results enable more precise modeling of the Galactic star formation rate encoded in the field WD mass distribution.\end{abstract}

\section{Introduction}
The initial-final mass relation (IFMR) is a mathematical mapping of initial mass $M_i$ of an intermediate mass star ($\approx 0.8-8M_\odot$) to the star's final mass $M_f$ at the end of its evolution when it has become a white dwarf (WD). Early studies of the initial and final masses of stars focused on constraining mass loss \citep{1977A&A....59..411W} and constraining the minimum zero age main sequence (ZAMS) mass at which a star will eventually undergo a core-collapse supernova \citep{1980ApJ...235..992R}. Subsequently, \citet{1983A&A...121...77W} were the first to recognize the value of establishing an IFMR. 

Much of the subsequent IFMR work has focused on covering the initial mass parameter space using WDs in open clusters \citep[e.g.,][]{1996A&A...313..810K,2000A&A...363..647W,2001ApJ...563..987C,2008ApJ...676..594K,2009ApJ...693..355W,Dobbie2009,2018ApJ...866...21C,2021A&A...645A..13P}, though highly useful constraints are also derived from binary field stars \citep[e.g.,][]{Liebert2005,2008A&A...477..213C,2015ApJ...815...63A}. A particularly novel approach to IFMR studies using Bayesian modeling of multiple open star clusters has been presented by \citet{2018MNRAS.480.1300S}; one advantage to their approach is that it considers the significant systematic impacts of cluster parameters more thoroughly than most other work has done.

Where an internally consistent set of WD and stellar evolutionary models has been employed, the dominant source of uncertainty in $M_i$ is the age of the open cluster \citep[e.g.,][]{Salaris2009}. However, there are many other possible sources of scatter in the relation including the effects of binary evolution \citep[e.g.,][]{2017PASA...34...58E,2020A&A...636A..31T}, magnetic fields, and variable mass loss due to angular momentum \citep{2019ApJ...871L..18C}. 

The open cluster IFMR has numerous constraints for initial masses $\gtrsim 2M_\odot$, yet the majority of field WDs used in studies of thin disk, thick disk, and halo stellar populations \citep[e.g.,][]{2017ApJ...837..162K,2019MNRAS.482..965K,2021arXiv210103341T} originated from stars with initial masses $\lesssim 2 M_\odot$ \citep[e.g.,][]{1987ApJ...315L..77W,2017ApJ...837..162K}.  A primary difficulty in detailing the low-mass end of the open cluster IFMR is the general lack of old ($\gtrsim 1$ Gyr) open star clusters due to their dissolution by Galactic tidal forces.  A related difficulty is that the WDs in a given old open cluster span a quite narrow range in $M_i$; WDs from more massive stars have cooled to the point where uncertainties in the WD cooling age make derivation of the progenitor nuclear lifetimes too uncertain for precise initial mass determination.  

Only a handful, though growing, number of open cluster IFMR constraints have been derived for WDs with progenitor lifetimes $\gtrsim 1$ Gyr.   \citet{2009ApJ...705..408K} present WD masses in the globular cluster \objectname{M4}, though their metal-poor progenitors may not be an appropriate match for metal rich field stars.  \citet{2008ApJ...676..594K} study the masses of WDs in the super-solar metallicity cluster \objectname{NGC 6791}, though the majority of these WDs are likely helium-core WDs which followed a non-canonical post main sequence evolutionary path.   Thorough analysis of WDs in the open clusters \objectname{NGC 7789} (age $\sim$ 1.5 Gyr; [Fe/H] $= -0.2$) and \objectname{NGC 6819} (age $\sim$ 2.4 Gyr; [Fe/H] $= -0.04$)  by \citet{2018ApJ...866...21C} results in a total of five likely single WDs from both clusters combined.  \edit1{Most recently, \citet{2020NatAs...4.1102M} analyzed the IFMR in light of new WDs in the old open clusters \objectname{Ruprecht 147} and \objectname{NGC 752} and improved WD parameter determination methodology to present evidence for a potential ``kink'' in the IFMR for progenitor star masses in the range of $1.65 M_\odot \lesssim M_i \lesssim 2.1 M_\odot$. With initial masses of $\approx 1.5M_\odot$, the WDs in M67 provide important constraints on this interpretation.}

In an effort to put further constraints on the low-mass end of the IFMR, we have obtained high signal-to-noise spectroscopy for a large sample of WDs in the old open cluster Messier 67.  \edit1{In our first paper on the M67 WD sample \citep[][hereafter Paper I]{2018ApJ...867...62W}, we focused on the ensemble properties of the WDs.  This included observational details, the WD sample selection, determination of WD cluster membership, the distribution of WD spectral types, the distribution of WD masses, potential WD remnants of blue stragglers, and a comparison of these properties to that of the WD field population.  In this paper we focus on the IFMR of the M67 WD sample and how it compares and contributes to the ever-growing semi-empirical open cluster IFMR.  To do so, we present details on the WD spectroscopic fitting technique used to derive WD masses and cooling ages; we discuss the careful pruning of the WD sample to include only well-measured objects that are highly likely to be cluster members that have experienced single star evolution; and we discuss the resulting measurements in light of recent developments in the open cluster IFMR.} Additional commentary concerning methodology, analysis, the IFMR, and caveats are presented in \citet{2018PhDT.......103C}.

\section{Observations and Spectral Analysis}
WD sample selection, spectroscopic observations and spectral extraction techniques are described in detail in \edit1{\citetalias{2018ApJ...867...62W}}.  
In short, spectra were obtained using the Low Resolution Imaging Spectrometer \citep{1995PASP..107..375O,1998SPIE.3355...81M} on the Keck I telescope. Observations employed multiple slitmasks with 1\arcsec\  wide slitlets and the 400 grooves mm$^{-1}$, 3400 \AA\ blazed grism.  We used the atmospheric dispersion corrector and so slitlets were not generally aligned with the parallactic angle.   The onedspec and twodspec IRAF packages were used to reduce the blue channel raw data\footnote{IRAF is distributed by the National Optical Astronomy Observatory, which is operated by the Association of Universities for Research in Astronomy (AURA) under a cooperative agreement with the National Science Foundation}.

The spectroscopic fitting routine described below requires model WD spectra to be convolved with the instrumental resolution.  Full-width at half-maximum (FWHM) resolutions are published on the Keck Observatory LRIS website\footnote{\url{https://www2.keck.hawaii.edu/inst/lris/dispersive_elements.html}} for each grism. As the spectral resolution varies slightly with slitlet placement and wavelength, we verified the  FWHM where possible using the measured FWHM of the  5577\AA\  auroral line of [\ion{O}{1}] \citep{1996PASP..108..277O}.  We adopt a resolution from the average of our measured FWHM and the published spectral resolution.   The adopted spectral resolutions are $\approx 8.0$~\AA. 
Empirically we find that the resulting WD physical parameters change by $\leq 0.5\sigma$ in $\log g$ if the assumed spectral resolution is changed by $\pm 1$\AA; the effect is insignificant for temperature determinations.  

\subsection{Spectroscopic Fitting Routine}
We derive physical parameters for each WD using the techniques and software described in \citet{Gianninas2011} and elaborated upon by \citet{2018PhDT.......103C}.  Briefly,  we compare our spectra with the models of \citet{Bergeron1992}, as updated by \citet{1995ApJ...449..258B}, \citet{2005ApJS..156...47L}, and \citet{Tremblay2009}. Using their spectral fitting routine we interpolate between models extending from 1,500 K to 140,000 K  in \teff and from 6.5 to 9.5 dex in \logg. We simultaneously fit the H Balmer series lines from H$\beta$ through H9 in each spectrum.

The routine convolves the model to the input spectral resolution for each observation. The observed and model spectra are then compared by calculating a $\chi^2$ figure of merit. A series of comparisons is performed over a grid of WD models varying in \teff and \logg with the Levenberg-Marquardt method \citep{1992nrfa.book.....P}, an algorithm based on the path of steepest descent, to determine the physical parameters that minimize the figure of merit. 

The spectroscopic fitting routine applies the mixing-length theory (ML2) of convection in one dimension for each object \citep{1992ApJ...387..288B}.
\citet{2011A&A...531L..19T} find temperature- and gravity-dependent corrections for 3D convective effects are necessary, especially for WDs with $\teff\leq 15,000$ K.  We therefore apply the ML2 correction functions of \citet{2013A&A...559A.104T} and use these corrected physical parameters throughout the rest of our analysis. 

As discussed in, e.g., \citet{Williams2007}, the formal fitting errors do not account for observational scatter and other external errors (especially flux calibration).  Following the discussion of \citet{2017ApJ...848...11B}, we have adopted the average external errors in Figure 8 of \citet{2005ApJS..156...47L} and add these in quadrature to the formal fitting errors to obtain our uncertainties in \teff and \logg.

We measure the signal-to-noise ratio (S/N) in the continuum between the H$\beta$ and H$\gamma$ lines using SPLOT in the onedspec package of IRAF. For each epoch, we use the appropriate dispersion solution and the adopted spectral resolution to convert the measured S/N per pixel to S/N per resolution element. 

\begin{figure*}
    \centering
	\includegraphics[width=0.9\columnwidth]{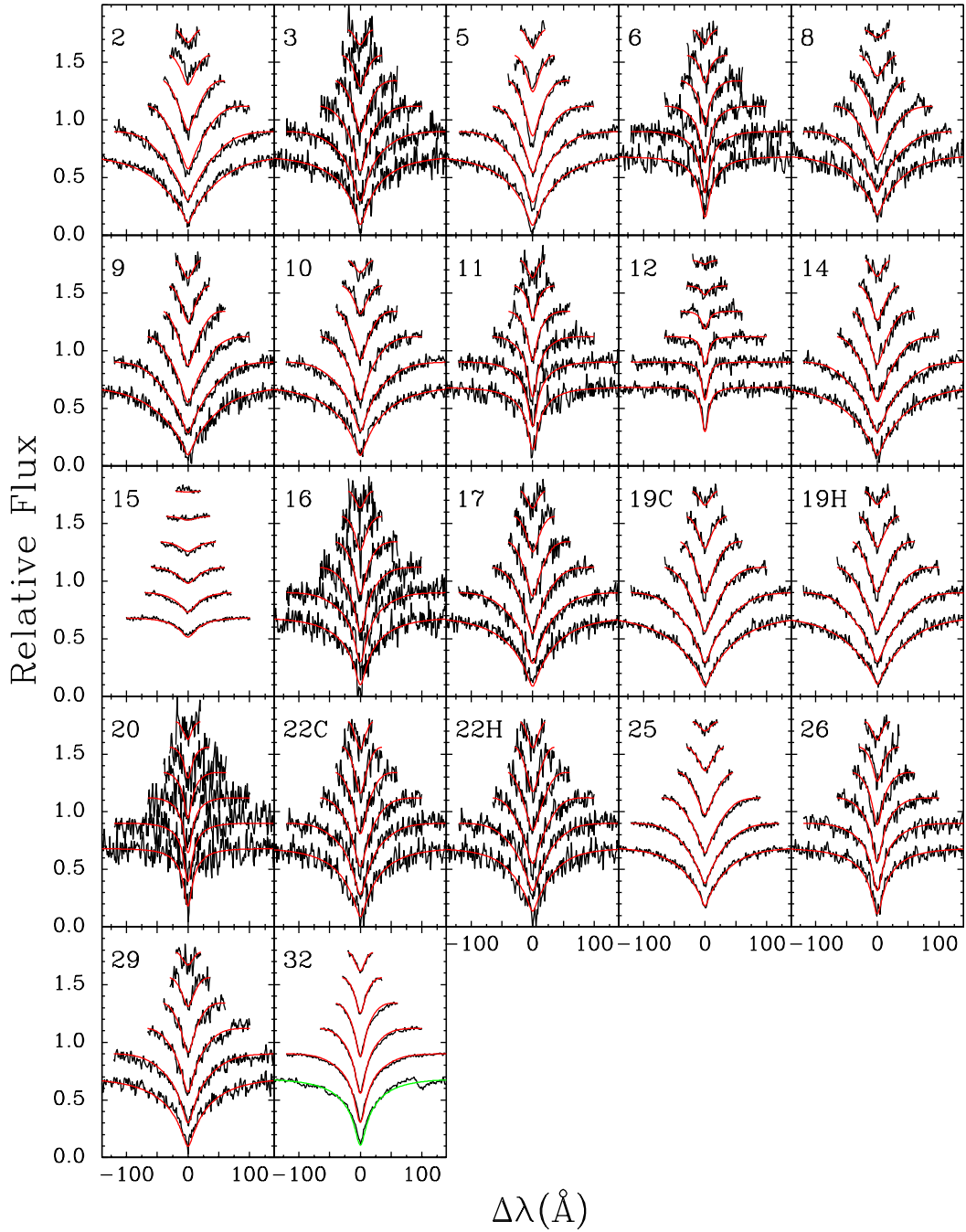}
    \caption{Normalized best-fit white dwarf models (red) overplotted on normalized Balmer-line profiles \edit1{(black)} for DA WDs in M67. Each panel shows the offset from line center in {\AA}ngstr{\"o}ms  plotted against normalized flux displayed from bottom (H$\beta$) to top (H9), with an arbitrary vertical offset in flux for clarity. \edit1{The green model for H$\beta$ in WD32 indicates that line was excluded from the fit due to contamination by a companion star (see text).} 
    \label{fig.M67NBL}}
\end{figure*}

Normalized Balmer line fitting often results in two local minima, with one typically at a higher $T_{\rm eff}$ and lower $\log g$ than the other.  We refer to these as the  ``hot" and ``cold" solutions. In many cases visual inspection of the Balmer line fits reveals the preferred solution.  In the cases where this visual inspection was not sufficient to identify the best solution, we compare the observed WD photometry \citepalias{2018ApJ...867...62W} to reddened model photometric indices derived from the calculations of \citet{2001PASP..113..409F,2006AJ....132.1221H,2006ApJ...651L.137K} and \citet{2011ApJ...730..128T} in order to select between the two solutions.  We present the physical parameters of our spectroscopic results in Table \ref{tab.physparams}, and we present our best fit model Balmer line profiles plotted over the observed WD profiles for each WD in Figure~\ref{fig.M67NBL}.

In the cases of M67:WD19 and M67:WD22, the photometry does not provide clear discrimination between the two solutions. We indicate the two possible solutions by appending H (for ``hot") and C (for ``cold") to the object names in Table~\ref{tab.physparams}.  We also exclude these WDs from the IFMR analysis that follows.

\begin{deluxetable*}{lccccccccccc}
\tablecaption{Physical Parameters for Messier 67 White Dwarfs \label{tab.physparams}}
\tablehead{\colhead{Star ID} & \colhead{RA} & \colhead{Dec} & \colhead{$T_\mathrm{eff}$} & \colhead{$dT_\mathrm{eff}$} & \colhead{$\log g$} & \colhead{$d\log g$} & \colhead{$M_f$} & \colhead{$dM_f$} &  \colhead{$\tau_\mathrm{cool}$} & \colhead{$d\tau_\mathrm{cool}$} & \colhead{S/N\tablenotemark{a}} \\ & J2000 & J2000 & K & K & & & $M_\sun$ & $M_\sun$ & log yr & log yr & }
\startdata
M67:WD2  & 8:50:39.43 & 11:53:26.82 & 14870 & 346 & 8.125 & 0.055 & 0.68 & 0.03 & 8.392 & 0.051 & 66\\
M67:WD3  & 8:50:47.60 & 11:43:30.04 & 10590 & 172 & 8.154 & 0.066 & 0.69 & 0.04 & 8.812 & 0.049 & 39\\
M67:WD5 & 8:50:52.52 & 11:52:06.82 & 11450 & 188 & 8.004 & 0.054 & 0.60 & 0.03 & 8.628 & 0.035 & 71\\
M67:WD6  & 8:50:58.61 & 11:45:38.79 &  8830 & 144 & 7.989 & 0.104 & 0.59 & 0.06 & 8.913 & 0.063 & 42\\
M67:WD8  & 8:51:01.78 & 11:52:34.48 & 20140 & 458 & 8.161 & 0.069 & 0.71 & 0.04 & 7.992 & 0.079 & 45\\
M67:WD9  & 8:51:05.31 & 11:43:56.66 & 13720 & 493 & 7.918 & 0.059 & 0.56 & 0.03 & 8.360 & 0.058 & 31\\
M67:WD10  & 8:51:08.89 & 11:45:44.87 & 11360 & 193 & 8.306 & 0.057 & 0.79 & 0.03 & 8.835 & 0.042 & 42\\
M67:WD11  & 8:51:09.14 & 11:45:20.21 &  9050 & 140 & 7.875 & 0.084 & 0.52 & 0.04 & 8.823 & 0.048 & 50\\
M67:WD12  & 8:51:09.58 & 11:43:52.63 &  7270 & 116 & 8.264 & 0.102 & 0.75 & 0.06 & 9.357 & 0.101 & 63\\
M67:WD14  & 8:51:12.11 & 11:52:31.32 &  13390 & 294 & 8.030 & 0.059 & 0.62 & 0.03 & 8.460 & 0.046 & 66\\
M67:WD15  & 8:51:19.90 & 11:48:40.63 &  53600 & 1041 & 7.629 & 0.066 & 0.54 & 0.02 & 6.241 & 0.042 & 55\\
M67:WD16  & 8:51:21.25 & 11:54:44.56 & 10030 & 163 & 8.035 & 0.083 & 0.62 & 0.04 & 8.798 & 0.052 & 34\\
M67:WD17  & 8:51:21.33 & 11:50:43.25 &  11360 & 201 & 8.036 & 0.060 & 0.62 & 0.03 &  8.656 & 0.041 & 50\\
M67:WD19C  & 8:51:24.92 & 11:53:56.61 &  13360 & 260 & 8.141 & 0.055 & 0.69 & 0.03 & 8.533 & 0.042 & 34\\
M67:WD19H  & 8:51:24.92 & 11:53:56.61 &  14950 & 289 & 8.076 & 0.051 & 0.65 & 0.03 & 8.352 & 0.045 & 34\\
M67:WD20  & 8:51:29.95 & 11:57:32.96 & 8450 & 149 & 7.687 & 0.141 & 0.42 & 0.06 & 8.809 & 0.062 & 39\\
M67:WD22H  & 8:51:37.75 & 11:58:43.22 & 15810 & 397 & 7.369 & 0.077 & 0.33 & 0.02 & 7.890 & 0.040 & 34\\
M67:WD22C  & 8:51:37.75 & 11:58:43.22 &  10460 & 187 & 7.578 & 0.085 & 0.39 & 0.03 & 8.522 & 0.038 & 34\\
M67:WD25  & 8:51:40.56 & 11:46:00.61 & 20440 & 308 & 7.977 & 0.045 & 0.61 & 0.02 & 7.768 & 0.057 & 68\\
M67:WD26  & 8:51:40.96 & 11:40:30.54 & 9580 & 146 & 7.934 & 0.072 & 0.56 & 0.04 & 8.792 & 0.043 & 68\\
M67:WD29  & 8:51:45.20 & 11:41:04.55 &  10830 & 178 & 8.252 & 0.062 & 0.75 & 0.04 & 8.854 & 0.047 & 81\\
M67:WD32  & 8:52:12.00 & 11:37:50.43 &  9780 & 139 & 7.933 & 0.048 & 0.56 & 0.02 & 8.768 & 0.031 & 220\\
\enddata

\tablenotetext{a}{Signal-to-noise per resolution element measured across the H$\beta$ and H$\gamma$ continuum (4500\AA\ to 4750\AA).}
\end{deluxetable*}

\subsection{Notes on Individual Objects}
The quality of our fits are generally excellent, but there are two primary exceptions.  We exclude the following objects from further analysis for these noted reasons:

\emph{M67:WD5} --- The quality of our fits for M67:WD5 are qualitatively poor despite the high signal-to-noise.  In particular, the best-fitting solutions fail to replicate the depth of the Balmer line cores. This object is not a proper-motion member of Messier 67 \citepalias{2018ApJ...867...62W}, so we engage in no further speculation.

\emph{M67:WD32} --- We identify M67:WD32 as a DA+dM system by the presence of the TiO absorption band near 4950~\AA\ in the red wing of H$\beta$.  We have excluded the blended H$\beta$ line from the fit and indicate this in Figure~\ref{fig.M67NBL} by coloring the excluded H$\beta$ line  in green rather than red. Although the resulting fit is qualitatively good, we exclude this object from the IFMR analysis because of the possibility of binary interaction in the past.

\section{The Cluster IFMR}

\subsection{Culling the White Dwarf Sample}
Construction and analysis of the semi-empirical IFMR assumes that each WD included is a coeval member of its respective open cluster and that each WD has experienced single-star evolution.  Here we discuss several methods we apply to the entire WD sample in order to maximize the likelihood that these assumptions are true.

\paragraph{Cluster Membership} -- Members of an open star cluster should have nearly identical parallaxes and proper motion vectors (technically they share the same convergent point, but for distant clusters this criterion asymptotes to common proper motions).   

\edit1{The precision proper motion and parallax measurements of the \emph{Gaia} early third data release \citep[\emph{Gaia} EDR3][]{2016A&A...595A...1G,2020arXiv201201533G} are available for six M67 WDs (M67:WD1, M67:WD12, M67:WD15, M67:WD21, M67:WD25, and M67:WD32), but are otherwise of limited utility for this study. } Due to their faint apparent magnitudes, most of the WDs in our M67 sample do not appear in the \emph{Gaia} EDR3 catalog.  We queried the \emph{Gaia} EDR3 catalog for our entire WD sample in Table 5 of \citetalias{2018ApJ...867...62W}, requiring positional matches within 1\farcs0 and \emph{Gaia} $G$ magnitudes within 0.5 mag of our published $g$ magnitudes; relaxing positional coincidence to 3\farcs0 did not change the results.  The matches and resulting parameters are given in Table~\ref{tab.gaia}.  

\begin{deluxetable*}{lcccccccl}
\tablecaption{Astrometric Data for M67 WDs in \emph{Gaia} EDR3 \label{tab.gaia}}
\tablehead{\colhead{Object}& \colhead{\emph{Gaia} Source ID}& \colhead{$\varpi$} & \colhead{$\mu_\mathrm{RA}$} &  \colhead{$\mu_\mathrm{Dec}$} & \colhead{$\Delta\mu / \sigma_\mu$} &  \colhead{PM Member?} & \colhead{Notes}  \\ & & mas & mas~yr$^{-1}$ & mas~yr$^{-1}$ & & & }
\startdata
WD1   & 604963941987041280 & $3.78\pm 1.86$  &  $-10.80\pm 1.63$  &   $-2.27\pm 1.46$  & 0.47 & yes & Spectral type DB\\
WD7   & 604914880575486336 & \nodata & \nodata & \nodata & \nodata & \nodata & Spectral type DB \\
WD12 & 604911238442984704 &  $6.96\pm 1.19$  &   $14.28\pm 1.52$  &  $-32.59\pm 0.93$ & 35.9 & no  & \\
WD14 & 604922164839680384 & \nodata & \nodata & \nodata & \nodata & \nodata & \\
WD15 & 604917698073661952 &  $1.20 \pm 0.25$ &  $-10.80 \pm 0.25$ &  $-2.70 \pm 0.17$ & 1.57 & yes & Photometric non-member\\
WD21 & 604898490980773888 &  $0.98 \pm 0.44$ &  $-11.48 \pm 0.43$ &  $-3.43 \pm 0.32$ & 1.94 & yes & Spectral type DB \\
WD25 & 604916353749379712 &  $1.41 \pm 1.20$ &  $-9.40 \pm 1.45$  &   $-2.65 \pm 1.50$ & 1.10 & yes & \\
WD30 & 604917079598643712 & \nodata & \nodata & \nodata & \nodata & \nodata & Spectral type DB \\
WD32 & 604899762291059712 &  $6.01 \pm 0.87$ & $-3.00 \pm  0.93$  & $-14.44 \pm 0.75$ & 17.5 & no  & DA+dM spectrum \\
\enddata
\end{deluxetable*}

The relevant astrometric parameters of Messier 67 as derived by \citet{2018A&A...616A..10G} are: $\varpi = 1.1325\pm 0.0011$ mas, $\mu_\mathrm{RA}=-10.9737\pm 0.0064$~mas~yr$^{-1}$, and $\mu_\mathrm{Dec} = -2.9396\pm 0.0063$~mas~yr$^{-1}$.   We determine if a WD is a proper motion member by calculating the difference between the WD and cluster proper motion vectors as follows:
$$ \left ( \frac{\Delta\mu}{\sigma_\mu} \right )^2 = \left ( \frac{\mu_\mathrm{RA,WD} - \mu_\mathrm{RA,M67}}{\sigma_{\mu_\mathrm{RA,WD}}} \right )^2 + \left ( \frac{\mu_\mathrm{Dec,WD} - \mu_\mathrm{Dec,M67}}{\sigma_{\mu_\mathrm{Dec,WD}}} \right )^2 $$
The result is equivalent to a standard deviation from the cluster proper motion vector if the uncertainties are both uncorrelated and follow a gaussian distribution, and if the cluster dispersion in both parameters is zero.  While none of these are technically valid assumptions, our interpretations are unlikely to change under a more rigorous analysis.  \edit1{The results, tabulated in Table \ref{tab.gaia}, are that M67:WD12 and M67:WD32 are excluded as a cluster member by proper motion.}  Further, as discussed in \citetalias{2018ApJ...867...62W}, M67:WD15 has an apparent distance modulus significantly larger than that of M67 despite its apparent astrometric membership, so it is also excluded from the IFMR analysis.

For the WDs lacking proper motion measurements in \emph{Gaia} EDR3, we use the proper motion memberships determined by \citet{2010A&A...513A..50B} to exclude M67:WD2, M67:WD5, and M67:WD8 as cluster members, and therefore we exclude them from the IFMR analysis as well.

\paragraph{Potential He-core WDs} -- Evolved stars with core masses $\lesssim 0.45\, M_{\odot}$ are unable to ignite He, resulting in a He-core WD.  Standard stellar evolution predicts that single stars with these low-mass cores have nuclear lifetimes significantly longer than a Hubble time, indicating that any such stars should not arise from single star evolution.   WDs with masses below the He-ignition threshold are generally thought to be the product of binary evolution \citep[e.g.,][]{1995MNRAS.275..828M,2011ApJ...730...67B}, though some argue that a metallicity-dependent high mass loss rate on the red giant branch could result in He-core WDs forming via single star evolution in the present-day universe \citep[e.g.,][]{2005ApJ...635..522H,Kalirai2007,2007ApJ...671..761K}.  Because of this uncertainty, we exclude the potential He-core WDs M67:WD20 and M67:WD22 from the IFMR sample.

\paragraph{Potential Blue Straggler Remants} -- We discuss in \citetalias{2018ApJ...867...62W} the possibility that M67:WD29 and M67:WD3 may be remnants of blue stragglers.  M67:WD29 is significantly more massive ($M_\mathrm{WD}=0.759\pm0.039 \msun$) than the mean mass of the cluster DAs ($\bar{M}=0.600\pm 0.010 \msun$), while M67:WD3 is mildly more massive than the mean ($M_\mathrm{WD}=0.693\pm 0.040 \msun$).  To be conservative, we exclude both WDs from the IFMR sample, though we note that the inclusion or exclusion of M67:WD3 does not change our quantitative conclusions significantly.

\edit1{\citet{2019ApJ...886...13J} claim that WD9 and WD25 are likely blue straggler progenitors based on their claim that  a 1.3\msun progenitor should create a 0.4\msun WD.  This claim is inconsistent with modern semi-empirical IFMRs \citep[e.g.,][]{2018ApJ...866...21C}, so we retain these two WDs in the sample. }

\paragraph{Overluminous WDs} -- Another indication of binarity could be if a WD appears overluminous for cluster membership.  While a WD in an unresolved binary may not have interacted with its companion, the analyzed spectrum will be a flux-weighted combination of the two stars' spectra, resulting in incorrect WD parameters.  We therefore exclude the candidate binary WDs M67:WD10 and M67:WD19 from the IFMR sample \citepalias{2018ApJ...867...62W}.

We admit that apparently single WDs in the M67 sample may have experienced binary interactions in the past \citep{2020A&A...636A..31T}.  This could include older double degenerate systems where a companion WD's luminosity has faded below detection limits, or WD+neutron star or WD+black hole systems that remained bound to the cluster after the supernova.  However, the tight distribution of DA WD masses \citepalias[$\sigma_{M_\mathrm{DA}}=0.043 \msun$,][]{2018ApJ...867...62W} suggests that the impact of any such binaries is negligible on our analysis.  

\paragraph{Remaining Sample} -- After the pruning discussed above, our final sample of DA WDs in M67 used in the IFMR analysis includes seven well-measured objects: M67:WD6, M67:WD9, M67:WD14, M67:WD16, M67:WD17, M67:WD25, M67:WD26.   

\subsection{Adopted Cluster Parameters}
As demonstrated by \citet{2018ApJ...866...21C}, when studying the IFMR it is crucial to select open cluster parameters such as distance, reddening, age, and metallicity in as consistent a manner as possible.  They also demonstrate how the use of a consistent set of stellar evolutionary models both in determining these cluster parameters and in calculating the WD initial masses greatly reduces systematic scatter in the resulting IFMR.  To simplify our comparisons with their IFMR and WD data, we focus on color-magnitude diagram derived parameters based on the PARSEC 1.2S isochrones \citep{2012MNRAS.427..127B,2015MNRAS.452.1068C}.

\citet{2015MNRAS.450.2500B} use the PARSEC evolutionary models and an iterative fit to the Hess diagram for M67 to derive optimum values of the cluster parameters.  Their best-fitting model has significantly subsolar metallicity, which is at odds with M67's spectroscopically determined near-solar metallicity \citep[e.g.,][]{2008A&A...489..677P}.  However, their second-ranked model more closely matches the spectroscopic metallicity, with an age of $3.542\pm 0.150$ Gyr, $Z=0.019\pm 0.003$, 
$E(\bv)=0.02\pm 0.02$, and a distance $d=880\pm 20$ pc.  These parameters are consistent with those used by \citet{2014MNRAS.444.2525C}, who also use the PARSEC models.  Even more compelling, the distance is fully consistent with the \emph{Gaia} DR2 parallax derived for the cluster by \citet{2018A&A...616A..10G}.  The  \citet{2015MNRAS.450.2500B} model-derived M67 distance corresponds to a parallax of $\varpi=1.136\pm 0.025$ mas, while the \emph{Gaia} DR2 cluster parallax is $\varpi = 1.1325\pm 0.0011$ mas. 

Adopting these values, we use the CMD 3.3 interface to the PARSEC version 1.2S tracks\footnote{\url{http://stev.oapd.inaf.it/cgi-bin/cmd}} to determine that the ZAMS mass of stars currently at the start of the thermally pulsing stage of the AGB is $M_i=1.438\pm 0.023\ \msun$.

\subsection{Calculating Initial Masses and Uncertainties}
We calculate the initial mass $M_i$ for each WD in our final sample using the methods from \citet{2009ApJ...693..355W}.  In summary, we interpolate the cooling time $\tau_{\mathrm{cool}}$ for each WD from the WD evolutionary sequences of \citet{2001PASP..113..409F}\footnote{Published at \url{http://www.astro.umontreal.ca/~bergeron/CoolingModels}} with thick H and He layers and uniformly mixed C/O cores; this cooling time and the associated uncertainty are tabulated in Table \ref{tab.physparams}.  This is subtracted from the cluster age to get the progenitor star's nuclear lifetime $\tau_{\mathrm{nuc}}$; this makes the reasonable assumption that the time each star spends in the thermally-pulsing AGB and post-AGB phases of evolution are negligible compared to $\tau_{\mathrm{cool}}$ and $\tau_{\mathrm{nuc}}$.  We then interpolate the PARSEC tracks to determine the ZAMS mass $M_i$ of a star with that $\tau_{\mathrm{nuc}}$.  As noted by \citet{Salaris2009}, there are numerous model-dependent systematics present in such an analysis, but these are dominated by the error in the assumed cluster age.

We determine the uncertainty in our initial and final masses via a Monte Carlo method.  We assume that the distribution of errors in the observed \teff and \logg of each WD are gaussian with a standard deviation given by the observational uncertainties quoted in Table \ref{tab.physparams}.  We then randomly draw 1000 \teff and \logg for each WD from this distribution and redetermine the $M_f$ and $M_i$ for each draw.  The 68\% envelope of the resulting distributions are tabulated as the uncertainties $dM_f$ and $dM_i$.   We do not consider the uncertainty in the cluster age at this point, because that will manifest itself as a systematic shift in all cluster WD $M_i$ values and not vary randomly from WD to WD.  Nonetheless, we calculate the systematic 68th percentile change in each WD's derived initial mass due to the cluster age uncertainties and present it separately with the stringent reminder that these are correlated systematic uncertainties for the entire ensemble.

\begin{deluxetable}{lccc} 
\tablecaption{Cluster White Dwarf Initial Masses \label{tab.initmass}}
\tablehead{\colhead{WD} & \colhead{$M_i$} & \colhead{$dM_{i,\mathrm{random}}$} & \colhead{$dM_{i,\mathrm{systematic}}$} \\
 & $(\msun)$ & $(\msun)$ & $(\msun)$}
\startdata
WD6  & 1.584 & $^{+0.029}_{-0.022}$ & $^{+0.034}_{-0.030}$ \\
WD9  & 1.483 & $^{+0.005}_{-0.004}$ & $^{+0.022}_{-0.022}$ \\
WD14 & 1.492 & $^{+0.005}_{-0.004}$ & $^{+0.023}_{-0.022}$ \\
WD16 & 1.547 & $^{+0.015}_{-0.012}$ & $^{+0.029}_{-0.025}$ \\
WD17 & 1.518 & $^{+0.007}_{-0.006}$ & $^{+0.025}_{-0.023}$ \\
WD25 & 1.459 & $^{+0.001}_{-0.001}$ & $^{+0.022}_{-0.022}$ \\
WD26 & 1.545 & $^{+0.012}_{-0.010}$ & $^{+0.029}_{-0.025}$ \\
\enddata
\tablecomments{Sharp-eyed readers may notice that these initial masses are systematically higher by a $\approx 0.03\msun$ than in Figure 9 of \citetalias{2018ApJ...867...62W}.  This is due to the masses in that figure having been calculated for $Z=0.0152$, inconsistent with our adopted metallicity of $Z=0.019$ from \citet{2015MNRAS.450.2500B}.  This error does not affect any of our conclusions from \citetalias{2018ApJ...867...62W}.}
\end{deluxetable}

The resulting $M_i$ values and 68\% uncertainty limits are given in Table \ref{tab.initmass}.  The remarkably tiny random uncertainties $dM_i$ for some cluster WDs are due to the very slow change in the main sequence turnoff mass as a function of time in old stellar populations such as M67 and the fact that $\tau_{\mathrm{cool}} \ll \tau_{\mathrm{nuc}}$.  For these WDs, the unaddressed systematics discussed by \citet{Salaris2009} are likely non-negligible compared to the stated uncertainties.  However, the errors in $M_i$ due to the uncertainty in the cluster age are still the overall dominant source for the cluster WDs.

As we illustrate in \citet{2020IAUS..357..179W} and \citet{2018PhDT.......103C}, the uncertainty distribution for individual WDs in the initial-final mass plane is usually strongly correlated and not gaussian.  Given the very narrow range in $M_i$ for the M67 WD sample, however, the correlation is dwarfed by the measurement and systematic uncertainties.  We therefore present the 68\% uncertainties in $M_i$ in Table \ref{tab.initmass} without considering the correlated errors.  We still emphasize the importance of the consideration of these correlated errors in any empirical IFMR studies. 

\section{Discussion}
\subsection{Comparison of M67 WDs to Existing IFMRs}

\begin{figure}
\includegraphics[width=0.95\columnwidth]{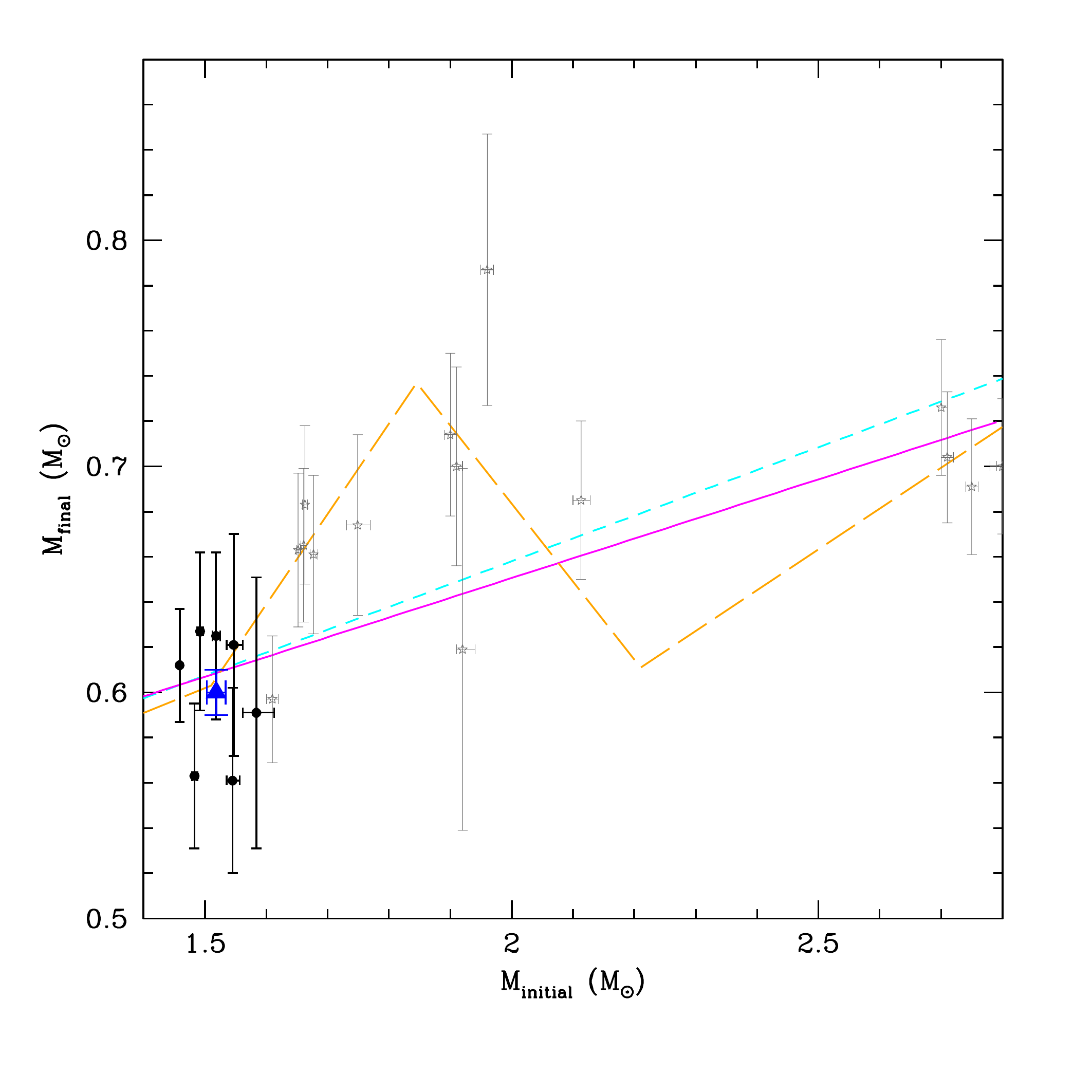}
\caption{Comparison of the M67 WD IFMR sample to two semi-empirical IFMRs for the region with $M_i$ between 1.4\msun and 2.8\msun.  Filled dark circles are the points for M67 WDs in the sample with error bars indicating the 68\% uncertainty range; the large blue triangle is the mean value for these points, with error bars indicating the errors on the mean.  The solid magenta line is the piecewise-continuous PARSEC IFMR from \citet{2018ApJ...866...21C}, \edit1{the long-dashed orange line is the ``kinked'' IFMR proposed in \citet{2020NatAs...4.1102M}, and the light grey stars with error bars are individual WDs from those publications.  The dashed cyan line is a linear fit to the IFMR from \citet{2018PhDT.......103C}. The M67 WDs are fully consistent with the functional fits from the earlier studies while providing a strong observational constraint for WDs with progenitor masses of $\approx 1.5M\sun$.}  \label{fig.ifmr_zoom}}
\end{figure}

\begin{figure}
\includegraphics[width=0.95\columnwidth]{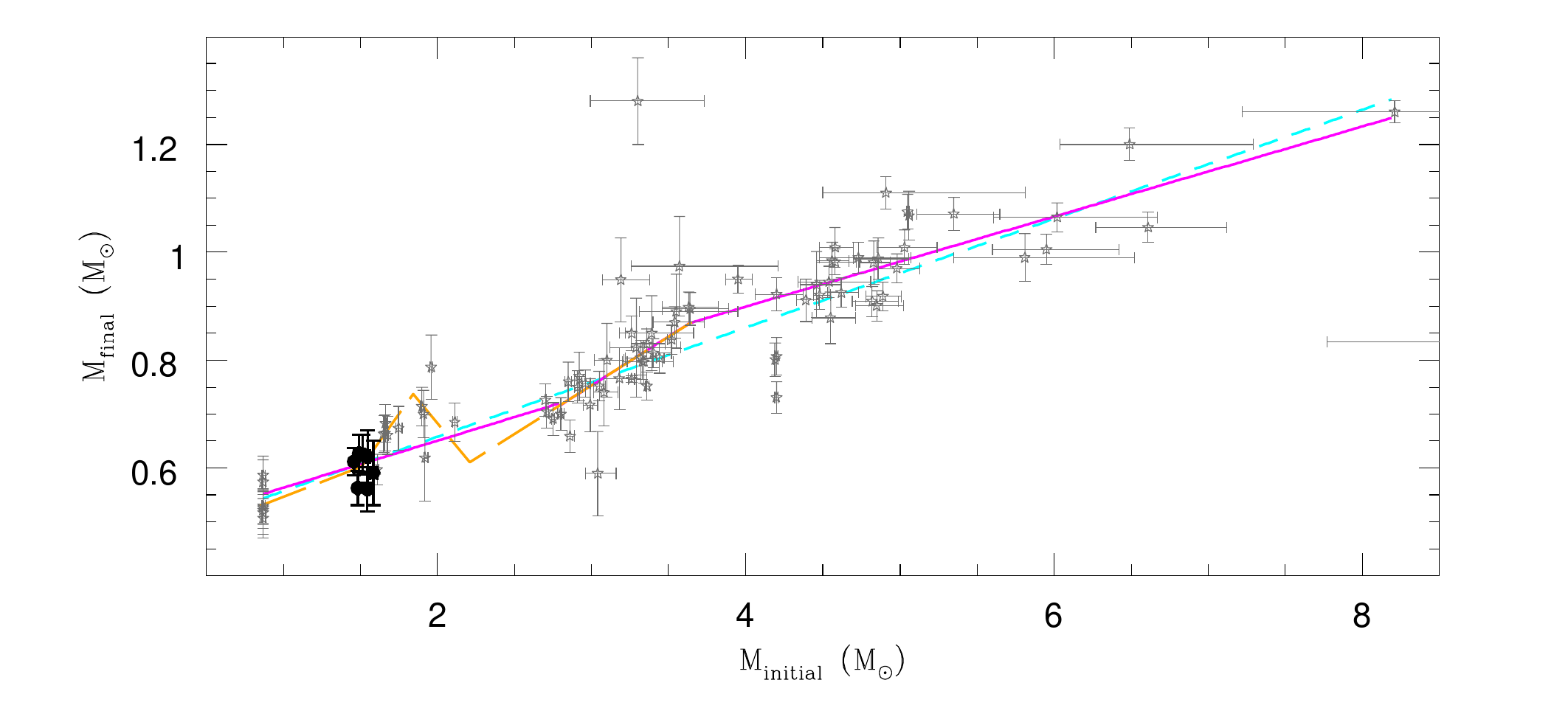}
\caption{Comparison of the M67 WD IFMR sample to the entire semi-empirical IFMRs of \citet{2018ApJ...866...21C}, \citet{2020NatAs...4.1102M}, and \citet{2018PhDT.......103C}.  Symbols are as in Figure \ref{fig.ifmr_zoom}.  Again, the M67 points are consistent with (within $1\sigma$ of) the existing semi-empirical IFMRs while providing crucial constraints for $M_{\mathrm{init}}\approx 1.5\msun$.  \label{fig.ifmr}}
\end{figure}

Of the multiple semi-empirical IFMRs that have been published, \edit1{the analyses of \citet{2018ApJ...866...21C} and \citet{2020NatAs...4.1102M} currently contain the largest sample of WDs analyzed in the most internally consistent manner currently possible.  As our WD sample was analyzed using nearly identical methods, we compare the M67 WD sample to these IFMRs.  We take the following steps in order to ensure we compare consistently derived values:  (1) our M67 WD analysis uses spectroscopically-derived WD parameters in combination with the PARSEC 1.2S stellar evolutionary models to derive initial masses; and (2) we compare the M6 IFMR points to the WD masses and initial masses in \citet{2020NatAs...4.1102M} derived from spectroscopic fits alone. We recognize that \citet{2020NatAs...4.1102M} show that a weighted average of spectroscopic \emph{and} photometric measurements of WD parameters result in less scatter in the semi-empirical IFMR, but we do not have photometric fits to the WD data available at this time.  Our conclusions are not affected by this choice.  }

\edit1{The comparisons are shown in Figure \ref{fig.ifmr_zoom}, which includes only a narrow range in $M_i$, and Figure \ref{fig.ifmr}, which displays the entire range of observed $M_i$.  The M67 data are highly consistent with both the \citet{2018ApJ...866...21C} and \citet{2020NatAs...4.1102M} piecewise linear IFMRs for initial masses of $\approx 1.5\msun$.}

In Figures \ref{fig.ifmr_zoom} and \ref{fig.ifmr} we also compare the M67 WDs to a linear fit of the IFMR from \citet{2018PhDT.......103C}.  This linear fit was generated from a reanalysis of the WD sample from \citet{2009ApJ...693..355W} using the \citet{Gianninas2011} methodology with the 3D corrections from \citet{2013A&A...552A..13T} and cluster ages rederived from PARSEC isochrones.  This fit also includes the M67 WDs from this work.  Finally, the linear fit also includes a rudimentary consideration of the correlated errors in $M_i$ and $M_f$ mentioned above.  \edit1{While this linear fit provides a decent qualitative fit to the semi-empirical IFMR over the entire range of progenitor masses, \citet{2020NatAs...4.1102M} find the reduced $\chi^2$ value of linear fits to be indicative of underfitting.}  Also qualitatively noticeable is the steeper slope of the data points in the $3~\msun$ to $4~\msun$ range as compared to a strictly linear fit, which inspired the piecewise-continuous IFMR of \citet{2018ApJ...866...21C}.

The M67 WDs also provide a useful sanity check for the low-mass end of the open-cluster IFMR, which lacks a significant number of points from solar-metallicity open clusters.  The WDs from \citet{2018ApJ...866...21C} with the lowest $M_i\approx 0.8\msun$ are from the globular cluster \object{Messier 4}, while all other WDs in the relation are from open clusters with near-solar metallicities. As progenitor star metallicity affects the resultant WD mass \citep[e.g.,][]{2015MNRAS.450.3708R}, concern about the effects of the introduction of significantly sub-solar metallicity stars in the open cluster IFMR seems warranted.  As the data from M67 show, any systematic effects introduced by the inclusion of Messier 4 WDs in the \citet{2018ApJ...866...21C} IFMR are limited to $M_i\lesssim 1.5\msun$.

\subsection{M67 WDs and the field WD population}

Since the M67 WDs in our sample arose from a small range of initial masses, and since published semi-empirical IFMRs have relatively shallow slopes over such a small range, \edit1{we can reduce the uncertainty in $M_i$ and $M_f$} by calculating the mean values these M67 WDs: $\bar{M}_i=1.518\pm 0.040\,\msun$, and $\bar{M}_f=0.600\pm 0.010\,\msun$, where the stated uncertainties are the formal errors on the mean, not the standard deviation of the sample.   \edit1{The complete M67 WD population undoubtedly contains higher mass WDs originating from higher progenitor masses, but these WDs have cooled below our spectroscopic magnitude limits given in \citealt{2018ApJ...867...62W}.}

This $\bar{M}_f$ is exceptionally close to the observed sharp peak of the field WD mass distribution of $M = 0.591 \msun,\,\sigma=0.035~\msun$ determined by \citet{2020ApJ...898...84K} from within the SDSS footprint \citep[see also][]{2019ApJ...871..169G,2018MNRAS.479L.113K,2016MNRAS.461.2100T,Gianninas2011,2010ApJ...712..585F}.   In addition to the extremely narrow peak at 0.59~\msun, the field WD mass distribution also shows an over-abundance of $0.7-0.9$ \msun WDs below 
$\teff=10,000$ K due to a delay in their cooling from the release of latent heat of  crystallization \edit1{and related effects \citep[e.g.][]{2020ApJ...902...93B,2020A&A...640L..11B,2020ApJ...902L..44C}.} Given the close agreement with the M67 WD masses, we conclude that the majority 
of local field WDs originated from $M=1.5~\msun$ main-sequence stars.

The field WD mass distribution encodes information about the star formation history of the Galactic disk; the mass distribution represents a convolution of the IFMR with the initial mass function and the star formation history of the disk population.  Therefore, the structure of the field WD mass distribution can provide independent constraints on the star formation history of the disk, if the IFMR is reasonably well-constrained.  \citet{2020ApJ...898...84K} compared the mass distribution for $\teff\geq 10,000$ K DA WD sample  against the predictions from a 10 Gyr old disk population with a constant star-formation rate, and found them to be consistent. However, they did not attempt to fit the mass distribution to derive the SFR for the disk. Several studies have suggested that the Milky Way thin disk experienced a burst of star formation $\sim 2-4$ Gyr ago \citep[e.g.,][]{2006A&A...459..783C,2013MNRAS.434.1549R,2018IAUS..334..158B,2019A&A...624L...1M}.  Some other recent studies of disk WDs have found weak evidence in support of this proposed star formation burst \citep{2019ApJ...887..148F,2019ApJ...878L..11I}.

As the age of M67 is consistent with the timing of this proposed burst in star formation, and as the M67 WD masses so closely match the mass peak of the local field WD population, it seems safe to assume that the peak of the field WD mass distribution is highly consistent with what would be expected for WDs whose progenitors formed during that peak of star formation, although detailed modeling is needed to confirm whether the broader WD mass distribution is consistent with such a peak.  More importantly, the tight constraints on the IFMR and WD evolution for $1.5~\msun$ stars should enable more precise modeling of the Galactic star formation rate encoded in the field WD mass distribution.

\acknowledgments
The authors wish to recognize and acknowledge the very significant cultural role and reverence that the summit of Maunakea has always had within the indigenous Hawaiian community.  We are most fortunate to have the opportunity to conduct observations from this mountain.  This material is based upon work supported by the National Science Foundation under grants awards AST-1910551, AST-1906379, and AST-0602288, and supported by a Cottrell College Science Award from the Research Corporation for Science Advancement.  The observing expertise of K.H.R.~Rubin was also highly valuable to the success of this project.  PC thanks A.~Gianninas for providing software training and technical support during this project. The authors appreciate valuable input provided by Matt Wood, Kent Montgomery, and the AAS Editors during the writing of this manuscript.  The authors also thank the anonymous reviewer for their time and effort.  This research has made use of NASA's Astrophysics Data System. This work has also made use of data from the European Space Agency (ESA) mission {\it Gaia} (\url{https://www.cosmos.esa.int/gaia}), processed by the {\it Gaia} Data Processing and Analysis Consortium (DPAC, \url{https://www.cosmos.esa.int/web/gaia/dpac/consortium}). Funding for the DPAC has been provided by national institutions, in particular the institutions participating in the {\it Gaia} Multilateral Agreement.

\facility{Keck:I (LRIS-B)}

\software{IRAF \citep{1986SPIE..627..733T,1993ASPC...52..173T}}


\end{document}